\documentstyle[a4,mhd209,cite,amssymb,graphicx,epsf]{article}

\title{Decay of metastable patterns for the Rosensweig instability:
revisiting the dispersion relation}

\author{Adrian Lange}

\institute{Universit\"at Magdeburg, Institut f\"ur Theoretische
Physik, Postfach 4120, D-39106 Magdeburg, Germany} 

\mhdhead{39}{1}{65}
\begin{document}
\maketitle

\begin{abstract}
The decay of metastable patterns in the form of magnetic liquid ridges
is studied in the frame of a linear stability analysis where the
ridges are the most unstable linear pattern of the Rosensweig
instability. The analysis of the corresponding dispersion relation
reveals that different sets of solutions exist. Two of them are
associated with an oscillatory and a pure exponential decay,
respectively.
\end{abstract}

\section{Introduction}
The most well known phenomenon of pattern formation in magnetic
fluids is the Rosensweig or normal field instability. This instability can
be observed for a horizontal layer of magnetic fluid (MF) with a free surface
with air above. Beyond a certain threshold $B_{\rm c}$ of a vertically
applied homogenous magnetic induction, the surface becomes unstable
and a stable pattern of peaks develops \cite{cowley67}. That final
pattern, resulting from nonlinear interactions, is one possible state
towards which the preceding most unstable linear pattern of liquid
ridges can develop if a supercritical induction $B_{\rm sup}\!>\!B_{\rm c}$
is abided \cite{lange01_wave}. The other possible state results from
the decay of the liquid ridges towards the flat surface
if the induction is switched from a supercritical to a subcritical value
$B_{\rm sub}$ \cite{reimann03}. With these two possible
final states, the most unstable linear pattern can be considered as a
metastable pattern formed by magnetic liquid ridges.

That ridges were studied in \cite{lange00_wave,reimann03} by means of
an experimental set-up in which the magnetic induction was {\it jumplike}
increased from a start value of $B_{\rm 0}<B_{\rm c}$ to $B_{\rm sup}$ \cite{lange00_wave}
and after a short time delay {\it jumplike}
decreased from $B_{\rm sup}$ to $B_{\rm sub}$\cite{reimann03}.
The jumplike change of the magnetic induction
allows to apply the linear stability theory, where the magnetic
induction is assumed to be instantly present. Consequently,
the quantitative agreement between measured and calculated data
is very good for the wave number of the linearly most unstable
ridges \cite{lange00_wave} and for the propagation velocity as well
as the oscillation frequency during the decay of the ridges
\cite{reimann03}. The calculation of the latter two quantities
disclosed new features of the dispersion relation which will be revisited
and analysed in the next section.

\section{Analysis of the dispersion relation}
A horizontally unbounded layer of an incompressible, nonconducting, and
viscous MF of infinite thickness and constant density $\rho$ is
considered. The MF has a free surface with air above, where the basic
state is that of a nondeformed surface. The dispersion relation for small
disturbances from the basic state is given by \cite{salin93}
\begin{equation}
\label{eq:1}
  \left( 1-{i\omega\over 2\nu q^2}\right)^2 +{1\over 4\rho \nu^2 q^4}
  \left[ \rho g q + \sigma q^3 - {(\mu_r -1)^2 B^2 q^2\over(\mu_r +1)
  \mu_0 \mu_r}\right]
  = \sqrt{ 1 -{i\omega \over \nu q^2} }\; ,
\end{equation}
where $g$ is the gravity acceleration, $\nu$ the kinematic viscosity of the MF,
$\sigma$ its surface tension with air, $\mu_r$ its relative
permeability, $B$ the absolute value of the external magnetic induction, and $\mu_0$ the
permeability of vacuum.
For the dispersion relation~(\ref{eq:1}), the small
disturbances were decomposed into normal modes, i.e. they are proportional to 
$\exp [-i(\omega\,t\!-\!{\bf q}\,{\bf r})]$. ${\bf r}\!=\!(x,y)$ is the planar space
vector, ${\bf q}\!=\!(q_x, q_y)$ the wave vector,
and $q\!=\!|{\bf q}|$ denotes the wave number. With $\omega\!=\!\omega_1 +i\omega_2$
the real part of $-i\omega$, $\omega_2$, is called the growth rate and defines
whether the disturbances will grow ($\omega_2 >0$) or decay ($\omega_2< 0$). The
absolute value of the imaginary part of $-i\omega$, $|\omega_1|$,
gives the angular frequency of the oscillation if $\omega_1$ is different from
zero. The critical induction for the onset of the Rosensweig instability at
$\omega =0$ is $B_c^2 = 2\mu_0 \mu_r (\mu_r +1)\sqrt{\rho \sigma g}(\mu_r -1)^{-2}$
\cite{cowley67}.

\begin{figure}[htbp]
  \begin{center}
    \includegraphics[scale=0.354]{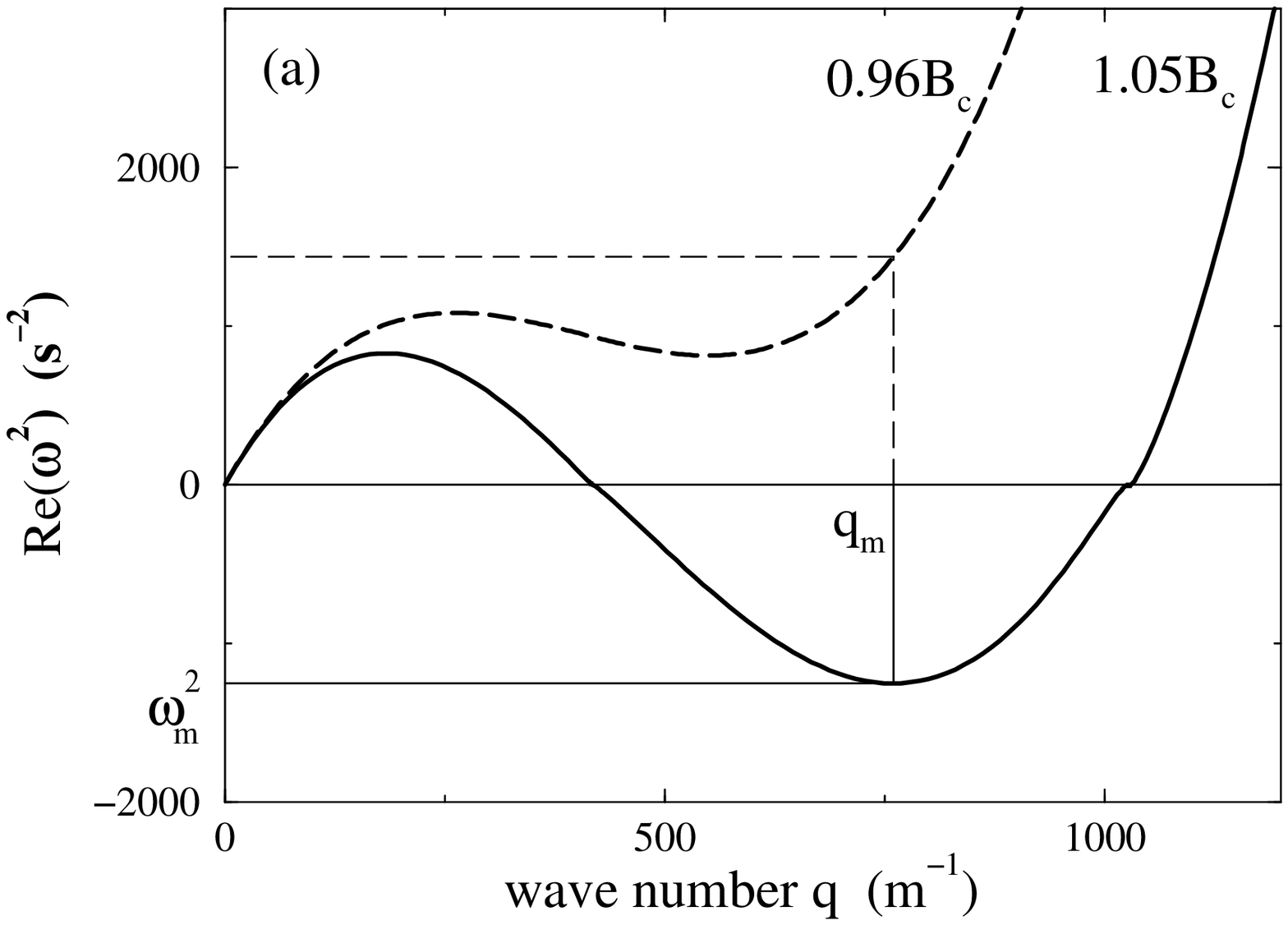}
    \hskip 0.2 cm
    \includegraphics[scale=0.354]{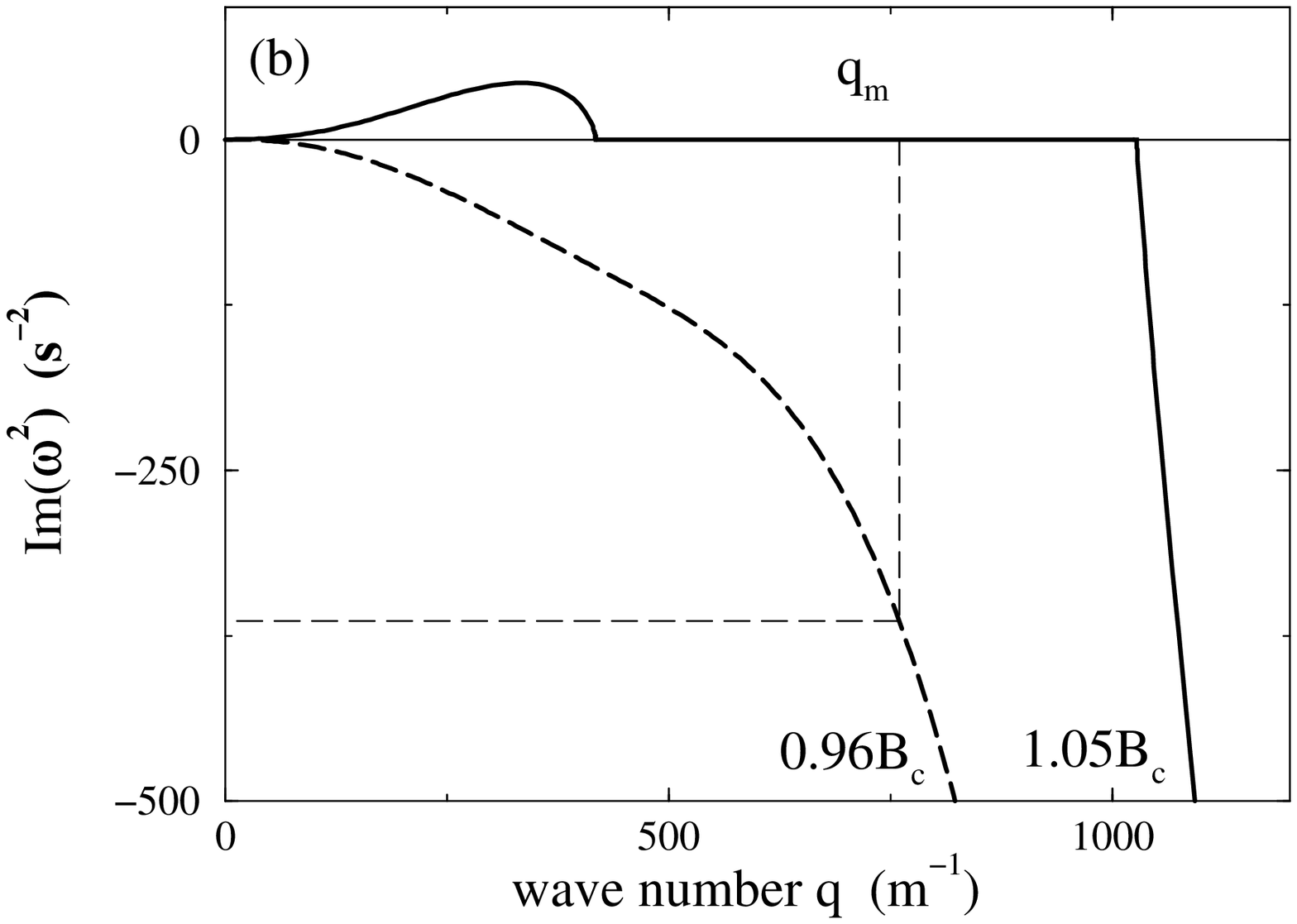}
    \caption{
    Dependence of ${\rm Re}(\omega^2)$ (a) and ${\rm Im}(\omega^2)$ (b)
    on the wave number $q$ for a supercritical
    ($1.05B_c$, thick solid line) and a subcritical ($0.96B_c$, thick dashed
    line) induction. The wave number $q_m$ of the
    most unstable linear pattern and the corresponding value $\omega_m^2$ are
    indicated by thin solid lines. The solution in the subcritical case
    for $q_m$ is indicated by thin dashed lines. The material parameters for
    the calculations are
    $\nu =5.17\cdot 10^{-6}$ m$^2$/s,
    $\rho = 1.16\cdot 10^3$ kg/m$^3$,
    $\sigma =2.65\cdot 10^{-2}$ kg/s$^2$,
    $\mu_r\simeq 1.935$, and $B_c=16.84$ mT.
    }
    \label{fig:disprel}
  \end{center}
\end{figure}

Figure~\ref{fig:disprel} shows for the reasons of clarity one
solution of the dispersion relation for
a supercritical induction $B_{{\rm sup}}=1.05 B_c$ (thick solid line) and for
a subcritical induction $B_{{\rm sub}}=0.96 B_c$ (thick dashed line). In the
supercritical case one can conclude from
${\rm Re}[\omega^2(q_{\rm m}, B_{{\rm sup}})]=\omega_{{\rm 1,m}}^2-\omega_{{\rm 2,m}}^2<0$ and
${\rm Im}[\omega^2(q_{\rm m}, B_{{\rm sup}})]=2\omega_{{\rm 1,m}}\,\omega_{{\rm 2,m}}=0$ that
a most unstable linear patterns evolves with the maximal growth rate
$\omega_{\rm m}=i\omega_{\rm 2,m}$ ($\omega_{\rm 2,m}>0$) and its
corresponding wave number $q_{\rm m}$ [thin solid lines in
Fig.~\ref{fig:disprel}(a)]. In the subcritical case the relations
${\rm Re}[\omega^2(q_{\rm m}, B_{{\rm sub}})]>0$ and
${\rm Im}[\omega^2(q_{\rm m}, B_{{\rm sub}})]<0$ [thin dashed lines in
Fig.~\ref{fig:disprel}(a)(b)] do not allow a simple conclusion on
the values of $\omega_{{\rm 1}}$ and $\omega_{{\rm 2}}$. Therefore
the classical presentation of $\omega^2$ versus $q$ chosen for the
analysis of inviscid MFs \cite{zelazo69,bashtovoi85,rosensweig87} is only of
limited use for the analysis of viscous MFs.

\begin{figure}[htbp]
  \begin{center}
    \includegraphics[scale=0.363]{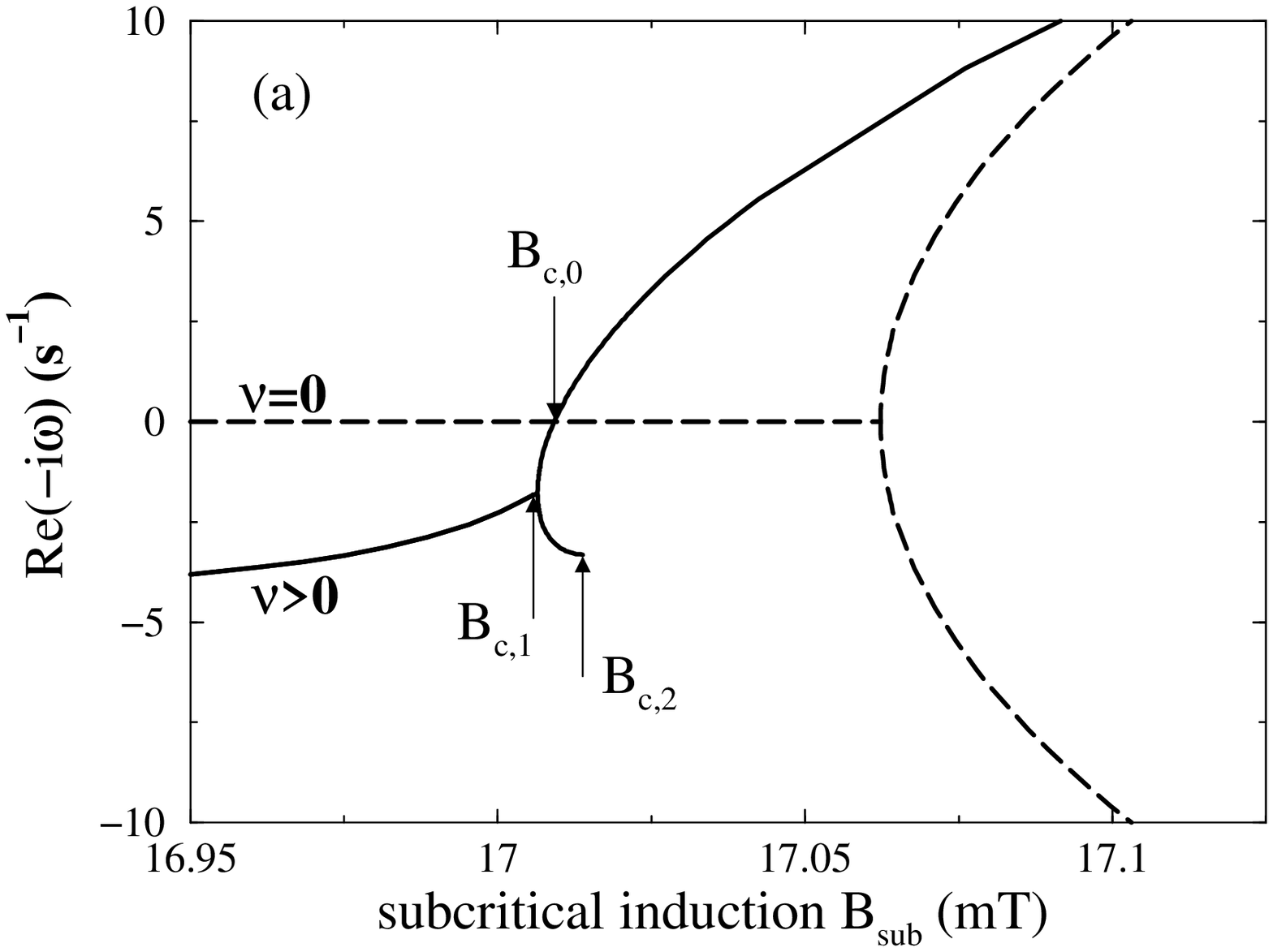}
    \hskip 0.2 cm
    \includegraphics[scale=0.363]{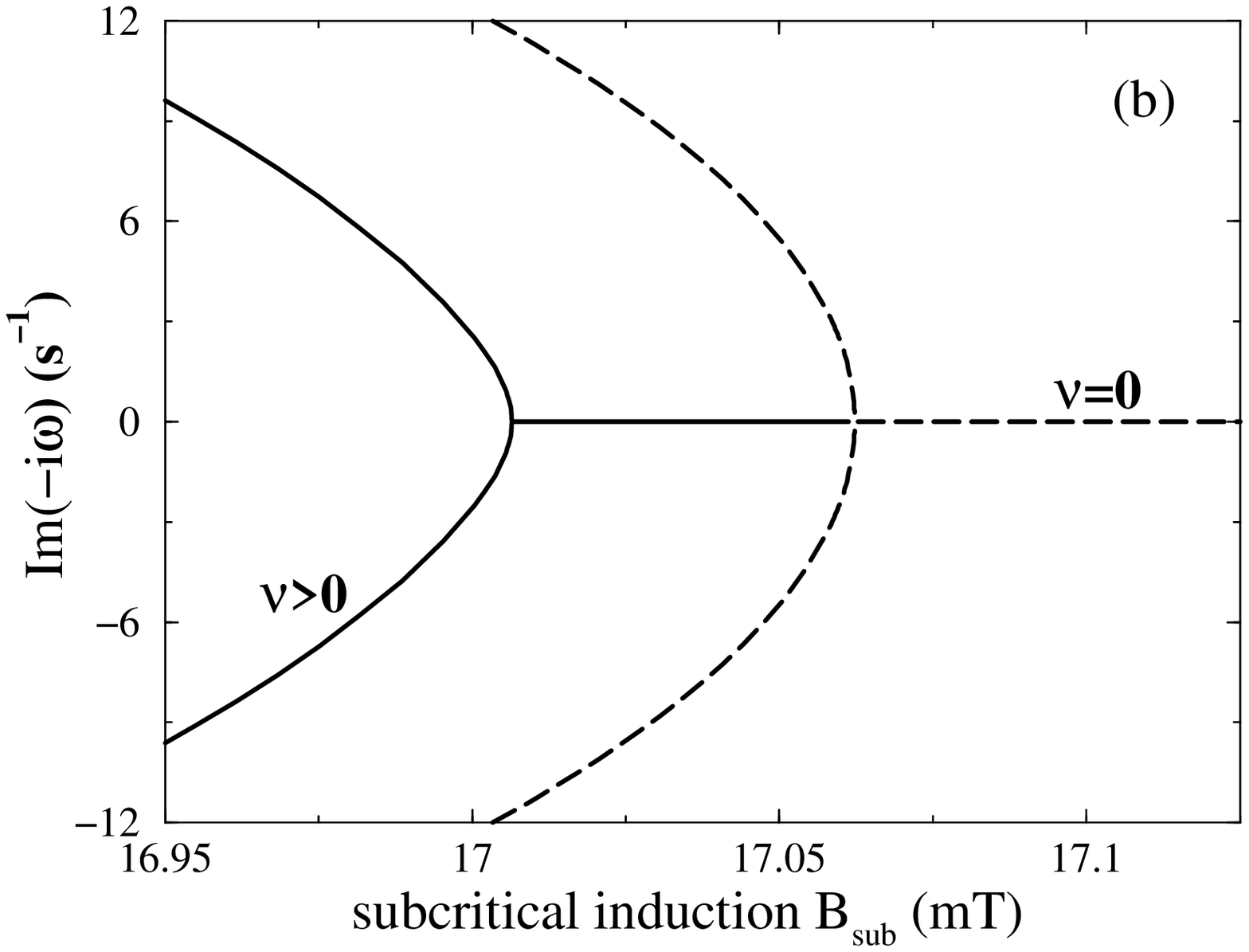}
    \caption{
    Dependence of ${\rm Re}(-i\omega)$ (a) and ${\rm Im}(-i\omega)$ (b)
    on the subcritical induction $B_{\rm sub}$ for an inviscid MF
    ($\nu =0$, dashed line) and a viscous MF ($\nu = 5.17\cdot 10^{-6}$ m$^2$/s, solid line). 
    For the former fluid two imaginary solutions are replaced by two
    real solutions with increasing $B_{\rm sub}$. For the latter fluid
    a region with two complex solutions is followed by a region with two real
    solutions which is succeeded by a region with one real solution.
    The three critical inductions $B_{\rm c,0}$, $B_{\rm c,1}$, and $B_{\rm c,2}$
    are explained in the text. The remaining material
    parameters for the calculations are those from Fig.~\ref{fig:disprel}
    with $q_{\rm m}(\nu )$ determined for $B_{\rm sup}=18$ mT.
    }
    \label{fig:real_imag_vs_B}
  \end{center}
\end{figure}

A more meaningful display is the plot of ${\rm Re}(-i\omega)$ and
${\rm Im}(-i\omega)$ versus the subcritical induction for different
viscosities (see Fig.~\ref{fig:real_imag_vs_B}), where $q_{\rm m}(\nu )$
of the most unstable linear pattern was determined for $B_{\rm sup}=18$
mT. If the fluid is inviscid (dashed lines in Fig.~\ref{fig:real_imag_vs_B}),
the pattern oscillates
below a certain threshold for the subcritical induction, $B_{\rm c,0}[q_{\rm
m}(\nu =0), \omega_2=0]$ [see Eq.~(\ref{eq:6})]. Above that threshold the
pattern can either decay according to the solution ${\rm Re}(-i\omega)<0$ or
can develop towards the most unstable linear pattern belonging
to $B_{\rm sub}\!>\!B_c$ since ${\rm Re}(-i\omega)>0$ is also a solution.

For a viscous fluid (solid lines in Fig.~\ref{fig:real_imag_vs_B})
the behaviour is more complex. A first critical induction, $B_{\rm c,1}$,
occurs, where the set of solutions for the dispersion relation~(\ref{eq:1})
changes from two complex solutions to two negative real
solutions. [The same transition occurs for the damping of gravity-capillary
waves as parameters as the aspect ratio and the so called inverse Reynolds
number are varied (see Figs.~4--6 and 8,9 in \cite{nicolas02}).]
Both real solutions exist until at a second critical induction,
$B_{\rm c,2}$, one of them abruptly ends. At a third critical induction,
$B_{\rm c,0}$, one of the two negative real solutions changes its sign and
becomes positive.

To understand this complex behaviour, the dispersion
relation~(\ref{eq:1}) in dimensionless units (indicated by a bar)
is analysed in the rearranged form
\begin{equation}
\label{eq:3}
\left( 1 + {-i\bar\omega\over 2\bar\nu \bar q\,^2}\right)^2-\sqrt{1 +
{-i\bar\omega\over \bar\nu \bar q\,^2}}
={-\bar q-\bar q\,^3+2\bar B^2\bar q^2\over 4\bar\nu^2 \bar q^4}\; .
\end{equation}
All lengths were scaled with $[\sigma/(\rho\, g)]^{1/2}$, the time with
$\sigma^{1/4}/(g^{3/4}\rho^{1/4})$, the viscosity with $\sigma^{3/4}/(g^{1/4}
\rho^{3/4})$,  and the induction with $B_c$.
Equation~(\ref{eq:3}) reveals that whatever value $\bar\omega\in{\Bbb C}$ has,
the left hand side of Eq.~(\ref{eq:3}) has to be real because the right hand side of
Eq.~(\ref{eq:3}) is {\it always} real since $(\bar q, \bar\nu, \bar B)\in{\Bbb
R}$. That condition together with the
mixing of real and complex quantities in Eq.~(\ref{eq:3}) is essential to
understand the above described appearance of different sets of solutions.
As long as $B_{\rm sub}\geq
B_{\rm c,1}$ all solutions of the dispersion relation with $q=q_m$ are
real, i.e. $-i\omega = \omega_2\in{\Bbb R}$ (see Fig.~\ref{fig:real_imag_vs_B}). Using
this result it follows from Eq.~(\ref{eq:3}) that there is a value 
$\bar\omega_2= -\bar\nu \bar q_{\rm m}^2$ beyond which the radicand becomes
negative. Since a complex value for the left hand side of Eq.~(\ref{eq:3}) is
not allowed, the solution does not exist beyond
$\bar\omega_2= -\bar\nu \bar q_{\rm m}^2$. This
corresponds to the point in Fig.~\ref{fig:real_imag_vs_B}, where one of the
solutions suddenly terminates at $B_{c,2}$. Therefore the second critical
induction yields to
\begin{equation}
\label{eq:4}
\bar B_{c,2}\bigr[\bar\nu, \bar q_{\rm m}(\bar \nu ),\bar\omega_2=-\bar\nu
\bar q_{\rm m}^2\bigr] =\sqrt{{1\over 2}\left( {1\over \bar q_{\rm m}}+
\bar q_{\rm m} +\bar\nu^2 \bar q_{\rm m}^2\right)}\; .
\end{equation}
The first critical induction is the minimal induction for which
real solutions exist, thus
\begin{equation}
\label{eq:5}
\bar B_{c,1}\bigr[\bar\nu, \bar q_{\rm m}(\bar \nu ),\bar\omega_{2,\min}\bigr]=
\sqrt{{1\over 2}\left( {1\over \bar q_{\rm m}}+\bar q_{\rm m}\right)-2\bar\nu^2 \bar q_{\rm 
m}^2
\left[ \sqrt{1+D} -\left( 1+{D\over 2}\right)^2\right] }
\end{equation}
with $\bar\omega_{2,\min}=\bar\nu \bar q_{\rm m}^2 \, D=(\bar\nu \bar q_{\rm 
m}^2/36)\bigr[-60+(108+12\sqrt{93})^{2/3}+144(108+12\sqrt{93})^{-2/3}\bigr]$.
Finally, the third critical induction is defined by $\bar\omega_2=0$ which leads to
\begin{equation}
\label{eq:6}
\bar B_{c,0}\bigr[\bar q_{\rm m}(\bar \nu ),\bar\omega_2=0\bigr]=\sqrt{{1\over 2}\left( 
{1\over
\bar q_{\rm m}}+\bar q_{\rm m}\right)}\; .
\end{equation}
At a closer inspection of Eqs.~(\ref{eq:4}-\ref{eq:6}) one realizes that the
three thresholds follow the relation $\bar B_{c,1}\leq \bar B_{c,0}\leq
\bar B_{c,2}$. Their dependence on the viscosity of the magnetic fluid will be
studied in the next section.

\section{Results and discussion}
\begin{figure}[htbp]
  \begin{center}
    \includegraphics[scale=0.5]{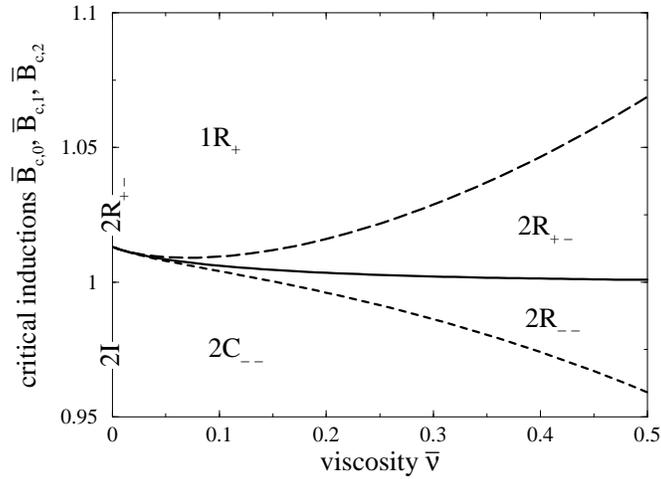}
    \caption{The three critical inductions $\bar B_{c,1}$ (dashed line),
    $\bar B_{c,0}$ (solid line), and $\bar B_{c,2}$ (long dashed line)
    versus the dimensionless viscosity $\bar\nu$. The regions with
    different sets of solutions for the dispersion relation are symbolised
    as follows: 2I -- two imaginary solutions, 2C$_{--}$ -- two complex
    solutions with negative real parts,
    2R$_{+-}$ -- two real
    solutions of different signs, 2R$_{--}$ -- two negative real solutions,
    1R$_{+}$ -- one positive real solution.}
    \label{fig:3kritB_vs_nu}
  \end{center}
\end{figure}

The behaviour of the three above defined thresholds on the viscosity is
studied, where all other material parameters are kept constant. The wave
number $q_{\rm m}(\nu)$ for the most unstable linear pattern is determined
for a supercritical induction of $18$ mT. Figure~\ref{fig:3kritB_vs_nu}
shows that for viscous MFs four regions with different sets of solutions for the
dispersion relation exist. For small subcritical inductions
the set of solutions consists of two complex solutions with negative real parts
(symbolised by 2C$_{--}$ in Fig.~\ref{fig:3kritB_vs_nu}) which is followed
by two negative real solutions (2R$_{--}$). Passing the threshold $\bar B_{c,0}$
(solid line), the set of solutions is formed
by one positive and one negative real solution (2R$_{+-}$). Only the
former solution lasts at high subcritical inductions (1R$_+$). From
Fig.~\ref{fig:3kritB_vs_nu} it becomes clear that for low viscous MFs,
$\bar\nu\leq 0.1$, only an oscillatory decay can be observed \cite{reimann03,comment}.
For high viscous MFs it
should be possible to observe an oscillatory (2C$_{--}$) as well as a pure
exponential decay (2R$_{--}$) of the pattern. Because for $\bar\nu\gtrsim 0.3$
the two thresholds $\bar B_{c,1}$ (dashed line)
and $\bar B_{c,0}$ are well separated and the
region 2R$_{--}$ becomes experimentally accessible.

For a nonmagnetic fluid
with the same density and surface tension, a dimensionless viscosity of
$\bar\nu\simeq 0.9$ is necessary to have a pure exponential decay, i.e.
the nonmagnetic fluid has to be considerable more viscous than the magnetic
fluid to result in the same behaviour. Another advantage in the use of a MF
is that the jumplike increase to $B_{\rm sup}$ is a simple way to prepare
just one particular metastable pattern whose decay can be studied.

For inviscid MFs Fig.~\ref{fig:3kritB_vs_nu} shows that only two different sets
of solutions occur. For subcritical inductions smaller than $\bar
B_{c,0}(\bar\nu =0)$ two pure imaginary solutions (2I) exist. Above that
threshold two real solutions with different signs (2R$_{+-}$) are found. 

\begin{figure}[htbp]
  \begin{center}
    \includegraphics[scale=0.354]{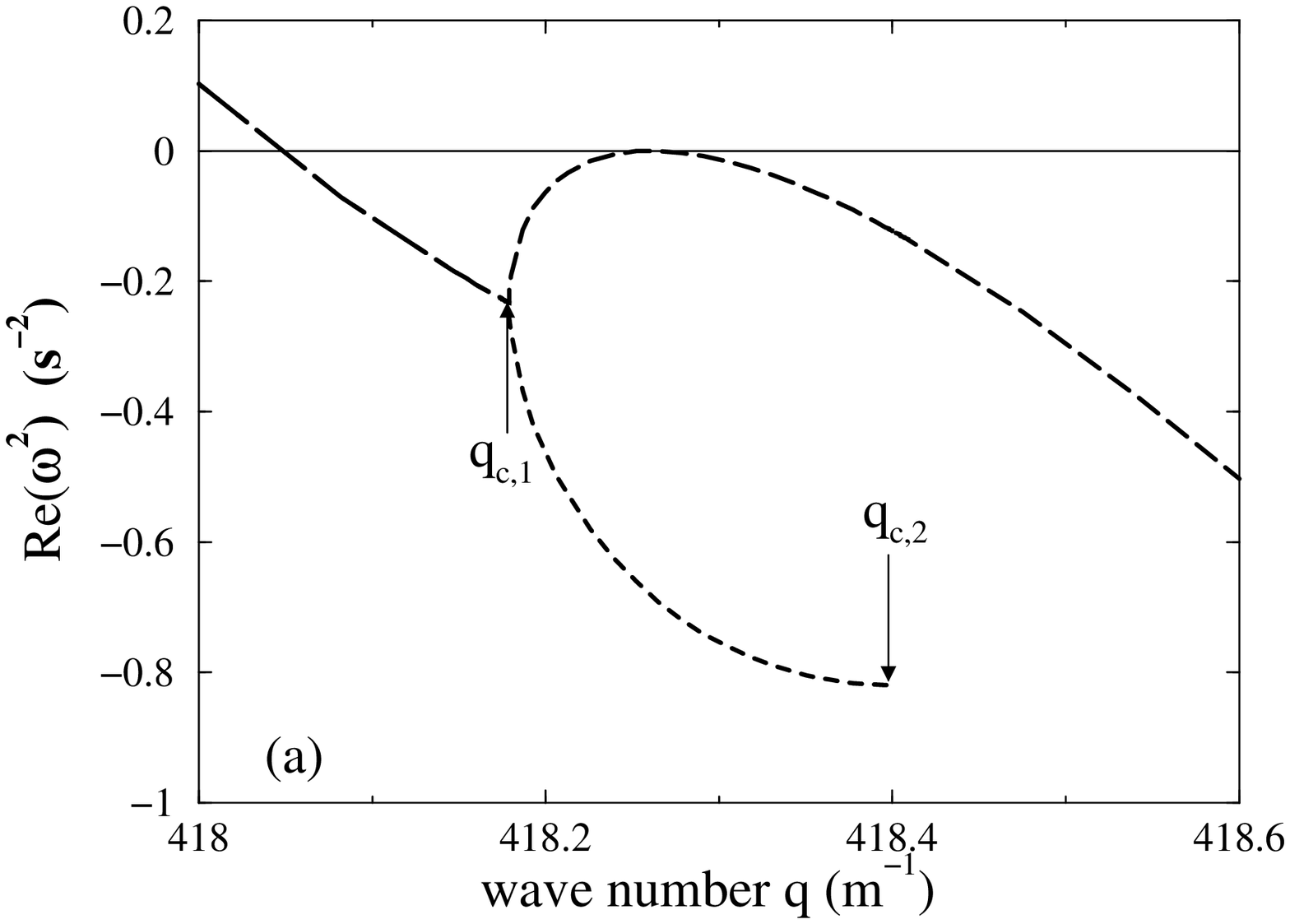}
    \hskip 0.2 cm
    \includegraphics[scale=0.354]{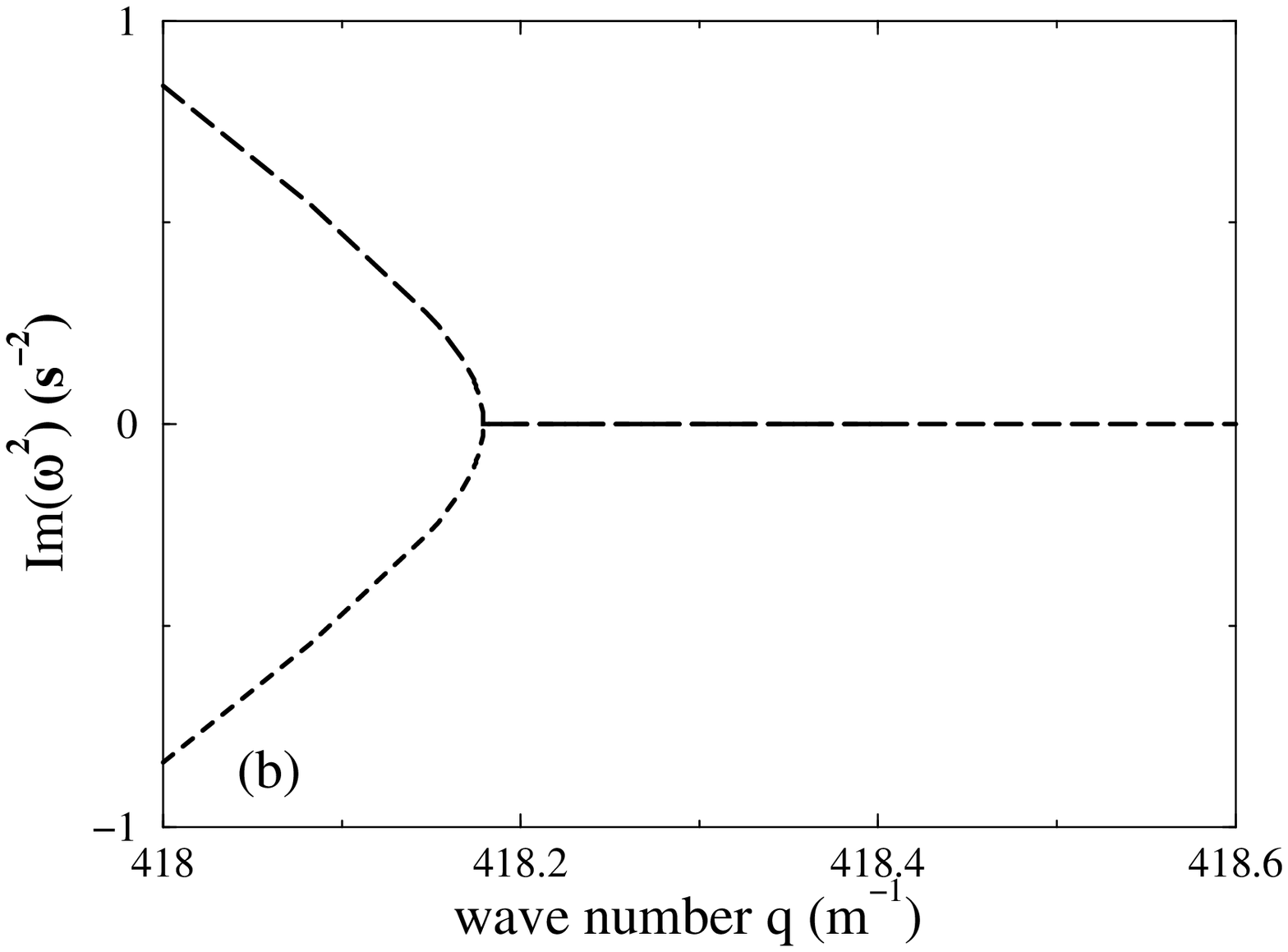}
    \includegraphics[scale=0.354]{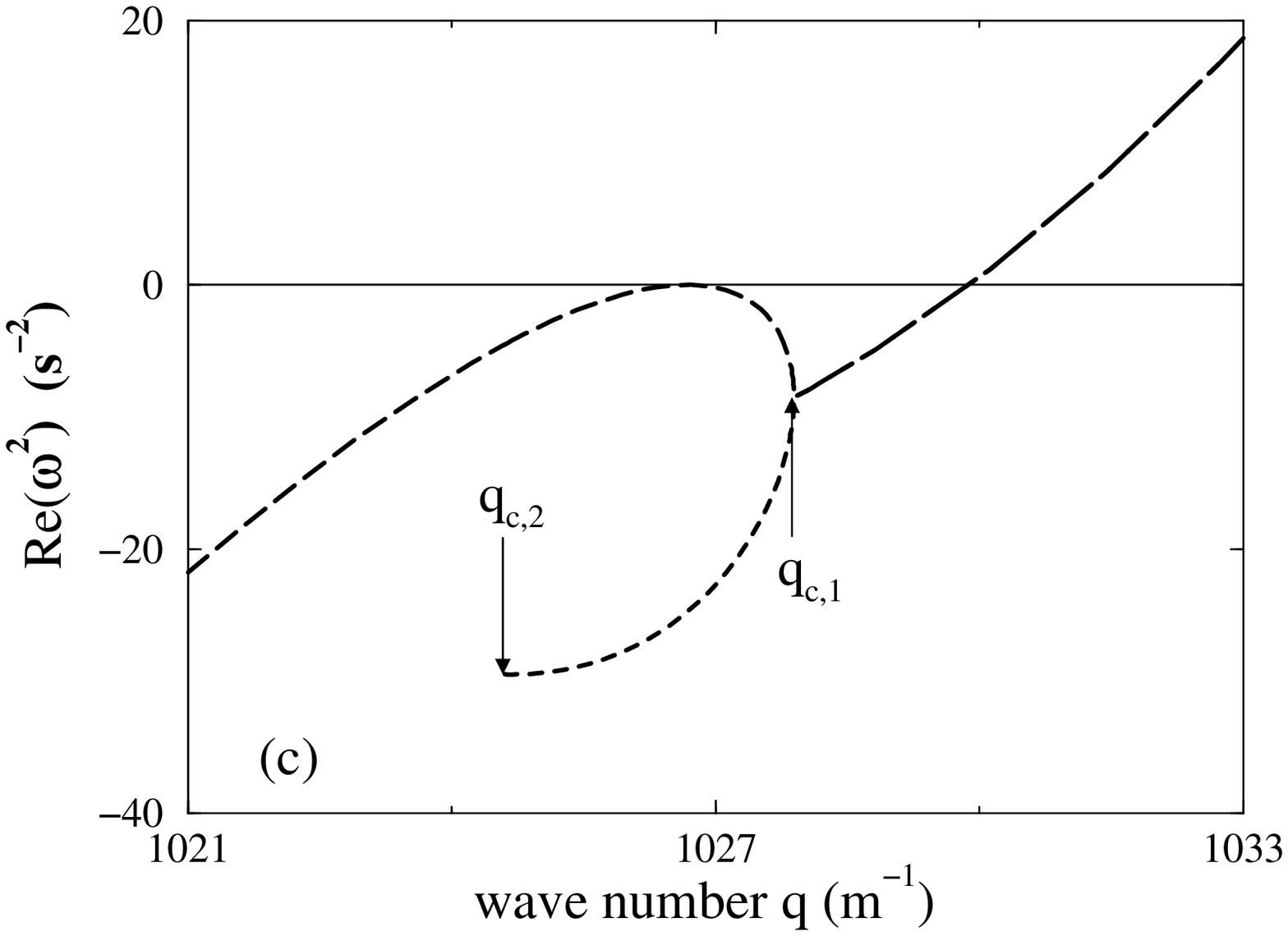}
    \hskip 0.2 cm
    \includegraphics[scale=0.354]{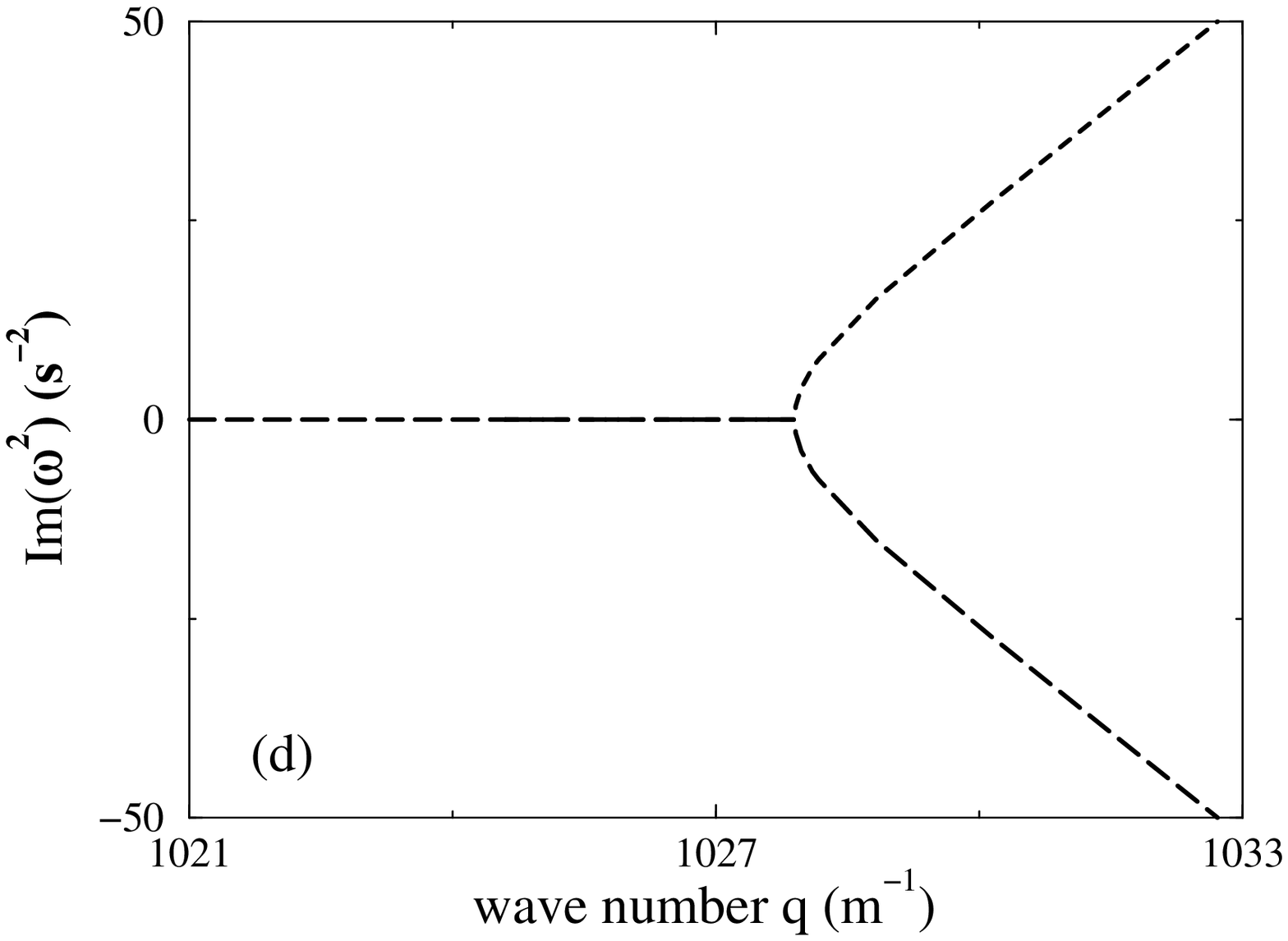}
    \caption{All solutions of Eq.~(\ref{eq:1}) for $B=1.05B_{\rm c}$
    in the form of
    ${\rm Re}(\omega^2)$ (a)(c) and ${\rm Im}(\omega^2)$ (b)(d) for
    wave numbers near $q\!\!=\!418.3$ m$^{-1}$ (a)(b) and
    $q\!=\!\!1027$ m$^{-1}$ (c)(d). In (a)(b) two complex solutions
    for $q\leq q_{\rm c,1}$
    are followed by two real solutions which exist for
    $q_{\rm c,1}\leq q\leq q_{\rm c,2}$ (dashed and long dashed lines).
    For $q_{\rm c,2}\!<\!q$ one real solution (long dashed line) exist.
    In (c)(d) the order of the different solutions is reverse: first
    one real solution, then two real solutions and finally two complex
    solutions.
    The material parameters are those from Fig.~\ref{fig:disprel}.
    Note the different scales at the axes.}
    \label{fig:zooms}
  \end{center}
\end{figure}

This analysis also leads to a better understanding in the case of a supercritical
induction, where regions with different sets of solutions should
exist, too. Whereas Fig.~\ref{fig:disprel} shows only one particular solution
of the dispersion relation~(\ref{eq:1}) for $0\leq q\leq 1200$ m$^{-1}$,
the two zooms around $q\!\!=\!418.3$ m$^{-1}$
[Fig.~\ref{fig:zooms}(a)(b)] and $q\!=\!\!1027$ m$^{-1}$
[Fig.~\ref{fig:zooms}(c)(d)] present all solutions. In contrast to the material
parameters for Fig.~\ref{fig:real_imag_vs_B}, now the induction is fixed at
$B_{\rm sup}\!=\!18$ mT and therefore transitions between different sets of solutions
occur at certain wave numbers. The
first zoom [Fig.~\ref{fig:zooms}(a)(b)] displays that for wave numbers below
$q_{\rm c,1}$ two complex solutions exist. Between $q_{\rm c,1}$ and $q_{\rm
c,2}$ two real solutions exist and beyond $q_{\rm c,2}$ only one real
solution. In a reverse order the different sets of solutions are appearing in the
second zoom [Fig.~\ref{fig:zooms}(c)(d)]. These features of the dispersion
relation have not been realized since only the solution for $q_{\rm m}$
was of interest in \cite{lange01_wave,lange00_wave,lange01_growth}. For that particular
wave number the  dispersion relation has only one real solution which
motivates the plot of one solution (thick solid line) in Fig.~\ref{fig:disprel}.

To conclude, in the frame of a linear stability analysis the decay of the
metastable pattern of magnetic liquid ridges is studied. There are initiated by a
jumplike increase of the magnetic induction to a supercritical value whereas
the decay is triggered by a jumplike decrease of the induction. The analysis
of the dispersion relation according to such a procedure reveals that
depending on the value of the subcritical induction different sets of
solutions exist (Fig.~\ref{fig:real_imag_vs_B}). The regions for these
different sets are separated by the three
thresholds~(\ref{eq:4}-\ref{eq:6}) whose dependence on the viscosity is
displayed in Fig.~\ref{fig:3kritB_vs_nu}. Whereas for low viscous MFs an
oscillatory decay of the ridges can be observed \cite{reimann03}, for
high viscous MFs a pure exponential decay should be detectable, too.

\section*{Acknowledgements}
The author benefited from helpful discussions with H. W. M\"uller, I. Rehberg,
B. Reimann, and R. Richter. This work was supported by the Deutsche
Forschungsgemeinschaft under Grant La 1182/2-2 and Ri 1054/1-2.

\received{13.12.2002}
\clearpage

%-------------------------------------Figures----------------------------------------------

\end{document}